\documentclass[letters,fleqn,usenatbib]{mnras}

\usepackage[T1]{fontenc}
\usepackage{ae,aecompl}
\usepackage[normalem]{ulem}

\usepackage{graphicx}	
\usepackage{amsmath}	
\usepackage{amssymb}	

\title[Uncovering the host galaxy of FBQS J1644$+$2619]{Uncovering the host galaxy of the $\gamma$-ray-emitting narrow-line Seyfert 1 galaxy FBQS J1644$+$2619}

\author[F. D'Ammando et al.]{F. D'Ammando$^{1,2}$\thanks{E-mail: dammando@ira.inaf.it}, J. A. Acosta-Pulido$^{3,4}$, A. Capetti$^{5}$, C. M. Raiteri$^{5}$, \newauthor R. D. Baldi$^{6}$, M. Orienti$^{2}$, C. Ramos Almeida$^{3,4}$ \\
$^{1}$Dipartimento di Fisica e Astronomia, Universit\`a di Bologna, Via Ranzani 1, I-40127 Bologna, Italy \\
$^{2}$INAF -- Istituto di Radioastronomia, Via Gobetti 101, I-40129 Bologna, Italy \\
$^{3}$Instituto de Astrofisica de Canarias, Calle Via Lactea, s/n, E-38205 La Laguna, Tenerife, Spain \\ 
$^{4}$Departamento de Astrofisica, Universidad de La Laguna, E-38205 La Laguna, Tenerife, Spain \\
$^{5}$INAF -- Osservatorio Astrofisico di Torino, via Osservatorio 20, I-10025, Pino Torinese, Italy \\
$^{6}$Department of Physics and Astronomy, University of Southampton, Southampton SO17 1BJ, UK
}

\date{Accepted 2017 March 20. Received 2017 March 20; in original form 2016 December 15}

\pubyear{2017}

\begin{document}
\label{firstpage}
\pagerange{\pageref{firstpage}--\pageref{lastpage}}

\maketitle

\begin{abstract}
The discovery of $\gamma$-ray emission from radio-loud narrow-line Seyfert 1
(NLSy1) galaxies has questioned the need for large black hole masses ($\ga
10^8$ M$_{\odot}$) to launch relativistic jets. We present near-infrared data
of the $\gamma$-ray-emitting NLSy1 FBQS J1644$+$2619 that were collected using
the camera CIRCE (Canarias InfraRed Camera Experiment) at the 10.4-m Gran
Telescopio Canarias to investigate the structural properties of its host
galaxy and to infer the black hole mass. The 2D surface brightness profile is
modelled by the combination of a nuclear and a bulge component with a S\'ersic
profile with index $n$ = 3.7, indicative of an elliptical galaxy. The
structural parameters of the host are consistent with the correlations of
effective radius and surface brightness against absolute magnitude measured
for elliptical galaxies. From the bulge luminosity, we estimated a black hole
mass of (2.1 $\pm$ 0.2)$\times 10^{8}$\,M$_{\odot}$, consistent with the values characterizing radio-loud active galactic nuclei.  
\end{abstract}




\section{Introduction}

The mechanisms and the physical parameters for producing a relativistic jet in radio-loud active galactic nuclei (AGN) are still unclear. One of
the key parameters seems to be the black hole (BH) mass, with only large
masses leading to an efficient jet formation \citep[e.g.][]{sikora07}. The
most powerful jets are found in luminous elliptical galaxies with very massive
BH ($M_{\rm\,BH}>10^{8}$--$10^{9}$ M$_{\odot}$). This is interpreted as indirect evidence that a high BH spin is required for the jet production. In
fact, elliptical galaxies should be the result of at least a major merger
event that is required for spinning up the central BH
\citep[e.g.][]{hopkins08}. It is becoming increasingly clear that radio-loud
AGN are the result of a specific path of galaxy evolution, involving mergers,
coalescence and spin-up of BH \citep[e.g.][]{capetti06, chiaberge15}. These mechanisms are expected to operate only at the high end of the galaxy mass distribution, and therefore in AGN hosted by elliptical galaxies. Only a handful of powerful radio galaxies have been found in spirals
  \citep[e.g.][]{morganti11,singh15}, all of them with BH mass $>$10$^{8}$ M$_{\odot}$.

The detection of variable $\gamma$-ray emission from a few radio-loud
narrow-line Seyfert 1 (NLSy1) by the {\em Fermi} satellite
\citep[e.g.][]{abdo09,dammando12,dammando15} challenges this view of the
mechanisms producing relativistic jets. NLSy1 represent a class of AGN
identified by their optical properties \citep{osterbrock85}: narrow permitted
lines [FWHM(H$\beta$) $<$ 2000 km s$^{-1}$, where FWHM stands for full width at half-maximum], [O$_{\rm\,III}$]/H$\beta$ $<$ 3, and a bump
due to Fe$_{\rm\,II}$ \citep[see e.g.][for a review]{pogge00}. It was suggested that
NLSy1 are young AGN with undermassive BH that are still growing
\citep[e.g.][]{mathur00}. Indeed, BH mass estimates are typically in the range $10^6$--$10^7$ M$_\odot$ \citep[e.g.][]{yuan08}. NLSy1 are generally radio-quiet, with only a small fraction \citep[$<$ 7$\%$,][]{komossa06} being radio-loud. Objects with very high values of
radio-loudness ($R > 100$) are even more sparse ($\sim$2.5\%). NLSy1 are usually hosted in late-type galaxies \citep[e.g.][]{deo06}, where the prevalence of pseudo-bulges are observed and the growth of the central BH is governed by secular processes rather than a major merger \citep[e.g.][]{mathur12}, although some objects are associated with early-type S0 galaxies \citep[e.g. Mrk 705 and Mrk 1239;][]{markarian89}. 

Luminosity, variability and spectral $\gamma$-ray properties of the NLSy1 detected by {\em Fermi}-LAT indicate a blazar-like behaviour \citep[e.g.][]{dammando16}. In addition, apparent superluminal jet components were observed in SBS 0846$+$513 \citep{dammando12}, PMN J0948$+$0022 and 1H
0323$+$342 \citep{lister16}. According to the current knowledge on how relativistic jets are generated and develop \citep[e.g.][]{boettcher02,marscher10}, these features should imply
$M_{\rm\,BH} > 10^{8}$ M$_\odot$ hosted in early-type galaxies. This seems to
disagree with the lower BH masses estimated for NLSy1. However, it was claimed
that the BH masses of these objects are underestimated due either to the
effect of radiation pressure \citep{marconi08} or to projection effects
\citep{baldi16}. As a consequence, NLSy1 might have larger BH masses, in
agreement with the values estimated by modelling the optical/ultraviolet
(optical/UV) data with a Shakura $\&$ Sunyaev disc spectrum
\citep[e.g.][]{calderone13}. Spiral galaxies are usually formed by minor
mergers, with BH masses typically in the range 10$^{6}$--10$^{7}$ M$_{\odot}$ \citep[e.g.][]{woo02}, so that it would not be clear how powerful relativistic jets could form in spiral galaxies. 

To obtain important insights into the onset of the production of relativistic jets in AGN it is crucial to determine the type of galaxy hosting
$\gamma$-ray-emitting NLSy1 and their BH mass. In this Letter, we report on the results of infrared (IR) observations of the NLSy1 FBQS J1644$+$2619 in the $J$ band collected using the Canarias InfraRed Camera Experiment (CIRCE) at the 10.4-m Gran Telescopio Canarias (GTC). FBQS J1644$+$2619 is an NLSy1 \citep[e.g.][]{yuan08} at redshift $z$ = 0.145 \citep{bade95} associated with a $\gamma$-ray source \citep{dammando15}. Here we perform a structural modelling of the FBQS J1644$+$2619 host galaxy and derive the BH mass estimate. Recently, this source has been observed in the $J$ and $K_s$ bands by \citet{olguin17} with the Nordic Optical Telescope (NOT). They decomposed the brightness profile up to $\sim$3.7 arcsec and modelled it with a pseudo-bulge, a disc and a ring component, with the addition of a stellar bar in the $K_s$ band alone. Our new $J$-band observations are deeper and extend the profile out to 5 arcsec, better constraining the host profile. This Letter is organized as follows. In Section 2, we report the CIRCE data analysis and results. In Section 3, we discuss the host galaxy morphology of FBQS J1644$+$2619, the BH mass estimate and compare them to what is known for the other $\gamma$-ray-emitting NLSy1. We draw our conclusions in Section 4. 

\section{Data Analysis}

\begin{figure}
\includegraphics[width=8.5cm]{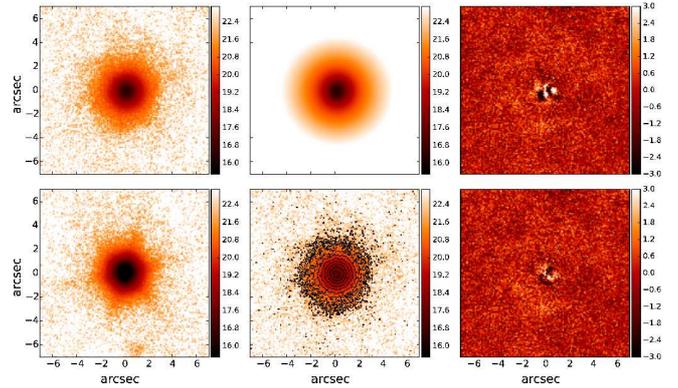}
\caption{2D surface brightness decomposition of FBQS J1644$+$2619. Top row: colour-scale $J$-band image of the galaxy (left-hand panel), the GALFIT
  model using S\'ersic+PSF (middle panel) and the residual image after subtracting the model (right-hand panel). Bottom row: the PSF1 (left-hand
  panel), the residual images after subtracting the scaled PSF1 (middle panel) and the GALFIT model using S\'ersic+PSF1 (right-hand panel). In all panels, north is up and east is right.}
\label{fig:galfmod}
\end{figure}

\subsection{Observations}

\begin{figure}
\includegraphics[width=8cm]{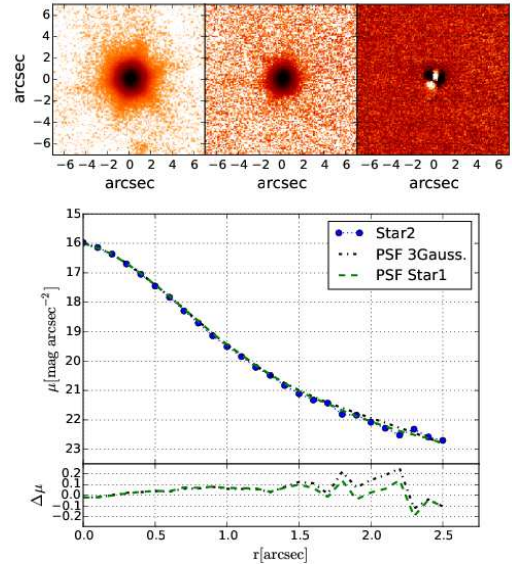}
\caption{As a test of the PSF, star 2 is fitted with the PSF1 and the three-Gaussian model. In the upper panel, from the left-hand to right-hand side: images of PSF1, star 2 and residuals after subtracting scaled PSF. The lower panel show the brightness profile of star 2, the three-Gaussian model and the scaled PSF1, with the residuals in the lower subpanel.}
\label{fig:testpsf}
\end{figure}

We obtained near-IR images in the $J$ band using the recently commissioned
instrument CIRCE \citep{Garner14} on the 10.4-m GTC. The instrument is
equipped with an engineering grade Hawaii2RG detector with a total field of
view of 3.4 $\times$ 3.4 arcmin$^2$ and a plate scale of 0.1 arcsec
pixel$^{-1}$. The observations were performed on 2016 July 16, in queue
mode. The seeing value was 0.9 arcsec. A predefined dither pattern of nine
positions was used. Frames were taken using individual integration times of
15 s (DIT = 15), in a Fowler 2 sampling readout mode, which was repeated three
times (NRAMPS = 3) at each dither position. The dither pattern was repeated seven times, to reach a total of 47 min on-target exposure time. 

Data reduction was performed using ad hoc routines written in IDL (Interactive
Data Language). The first step in the data processing includes the subtraction
of dark current frames. An illumination correction was obtained from twilight
sky exposures  in order to compensate  a decrease of about 40 per cent, from
the center to border of the field of view (FOV). A correction to remove a pattern of inclined stripes related to reading amplifiers was introduced at this point. Once this pattern was removed, the images corresponding to each dither cycle were median combined to form a sky frame, which was subtracted to each frame of the cycle. Finally, all sky-subtracted images are combined by the commonly used shift-and-add technique. During the combination of these frames, a bad-pixel mask was used, which includes the two vertical bands corresponding to non-functional amplifiers.

The photometric calibration was obtained relative to a nearby Two-Micron All-Sky Survey (2MASS) field star, 2MASS J16444411$+$2619043, with quality
flags `A' in all bands, with an error of 0.031 mag. The derived zero-point in
the $J$ band was 24.21 (flux in ADU s$^{-1}$), with an uncertainty of about
0.1, which was estimated by the comparison with other 2MASS field
stars. Sources with $J \sim 20.5$ are detected with signal-to-noise ratio $\sim 10$.

\begin{figure}
\includegraphics[width=8cm]{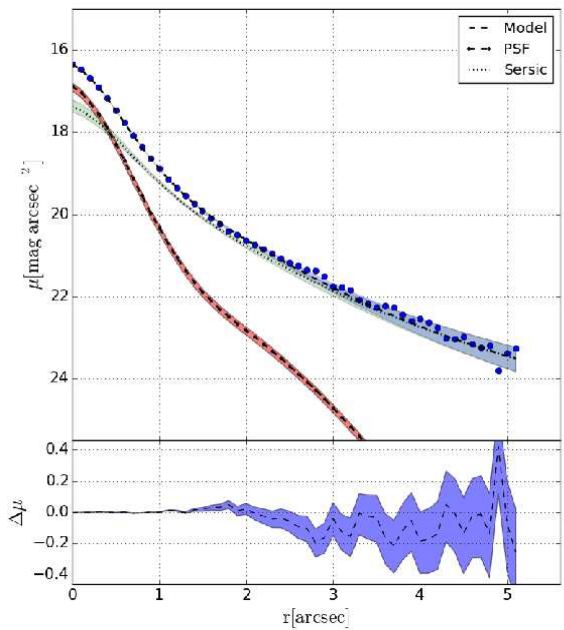}
\caption{Radial profile of the surface brightness distribution of FBQS
  J1644$+$2619 (upper panel) and residuals (bottom panel). The filled circles
  show the observations, and the black solid line, the red dot--dashed line and the green dashed line represent the  model, the PSF and the S\'ersic profiles, respectively.}
\label{fig:radprof}
\end{figure}

\subsection{Host galaxy structure}

In Fig.~\ref{fig:galfmod} (top left-hand panel), we show the central 12 $\times$ 12 arcsec$^{2}$ of the FBQS J1644$+$2619 deep $J$-band image. The
object appears resolved by visual inspection, but to study the structure more quantitatively, we used the 2D surface brightness model fitting
code GALFIT \citep{Peng02}. We considered a model consisting of a single bulge
component, described by a S\'ersic profile, and a point spread function (PSF)
to account for the unresolved nuclear component. The PSF was obtained from a
single nearby bright star ($J$ = 14.2), hereafter PSF1, at a distance of 25
arcsec east of the target. We could not consider other close and bright stars
to model the PSF due to the small FOV of our image. A PSF model was built
using three Gaussians of different widths. We tested the PSF model with a
different star in the field and the fit was successful showing no residual
excess (Fig.~\ref{fig:testpsf}). All parameters of the S\'ersic profile and
the scaling factor of the PSF were allowed to vary freely. The residuals of
the model PSF+S\'ersic (see Fig.~\ref{fig:galfmod}, upper panels) show a
structure in the centre ($r \simeq 1$ arcsec), due to details not included in
our PSF model. We also tried a model by considering PSF1 for the nuclear
component, and the residuals are largely reduced (see Fig.~\ref{fig:galfmod},
right-hand panels). Therefore, we assumed this model as the reference for the
structural galaxy host parameters. The best-fitting values are provided in
Table~\ref{ta:galfpar}. As a sanity check, we tried a model fitting after
fixing S\'ersic index to values $n$ = 1, 2, but the residuals are much worse
than for $n > 3$. The background was fixed to zero, which comes naturally from
the data processing, although it exhibits large local variations. We measured
the sky at different regions and obtained a 1$\sigma$ rms value, which was
used as an input for new GALFIT models. The resulting parameters are reported in Table~\ref{ta:galfpar} and are taken as the parameter uncertainty estimation. The residual image after subtracting only the scaled PSF1 shows a rather smooth structure, similar to an elliptical morphology (see Fig.~\ref{fig:galfmod}, lower panels). In addition, a positive residual region is observed around 3\, arcsec south. This brightness enhancement has been reported by \citet{olguin17}, suggesting that it may be the hint of a merger event.

The 1D brightness profile of FBQS~J1644$+$2619 is shown in Fig.~\ref{fig:radprof}, as well as the profiles of the PSF component and the
S\'ersic model. The profile extends out to a radius $r \sim$ 5 arcsec. The nuclear contribution becomes weaker than the galactic one already at $r < 1 \rm \,arcsec$ and this allows us to study the host in great detail.  

\begin{table*}
\caption{Photometric and structural parameters of the FBQS J1644$+$2619 host galaxy. The superscript and subscript values correspond to a sky value 1$\sigma$ above and below 0. Absolute magnitudes are computed using a distance module equal to 39.10, ($D_{L}=661\ {\rm Mpc}$) and a Galactic extinction correction $A_J=0.058$.}
\label{ta:galfpar}
\resizebox{\textwidth}{!}{\begin{tabular}{llllccccccc}
\hline
\multicolumn{4}{c}{PSF} & \multicolumn{6}{c}{Bulge - S\'ersic model} & \multicolumn{1}{c}{$\chi^{2}$$_{\nu}$} \\
PSF model & $m_J$ & $M_J$  & FWHM      & $m_J$ & $M_J$ & $n$ & $R_e$ & $b/a$ &  <$\mu_e$>  &  \\
     &  &       &   (arcsec) &      &       &   & (arcsec) / (kpc) &  &  (mag arcsec$^{-2}$)  & \\
\hline
Star 1 & $16.55^{16.48}_{16.72}$ &   -22.55   & 0.90  & $15.87^{15.99}_{15.71}$ & -23.23 & $3.7^{2.9}_{5.1}$ & $0.93^{0.88}_{1.01}$ / 2.29 & $0.934^{0.934}_{0.934}$ &  $17.71^{17.72}_{17.74}$ &  0.807$^{0.819}_{0.801}$ \\
Three Gaussians & $16.58^{16.50}_{16.70}$ &  -22.52    & 0.90  & $15.83^{15.96}_{15.62}$ & -23.72 & $4.1^{3.2}_{5.8}$ & $0.98^{0.91}_{1.11}$ / 2.29 & $0.960^{0.958}_{0.961}$ & $17.71^{17.72}_{17.74}$ & 0.865$^{0.876}_{0.859}$ \\
\hline
\end{tabular}}
\end{table*}

The best-fitting S\'ersic index is $n=3.7$, which is a good description of an elliptical galaxy \citep{blanton03}. The ellipticity obtained from the 2D modelling is $\epsilon = 1 - b/a = 0.066$, indicating that the host of FBQS J1644$+$2619 is an E1 type galaxy. The structural parameters of the host are consistent with the correlations of effective radius and surface brightness with absolute magnitude measured for elliptical galaxies in the Sloan Digital Sky Survey by \citet{bernardi03}. For consistency with their study, we repeated the fit with $n=4$, corresponding to a de Vaucouleurs profile. We obtained $R_{\rm e}$ = 0.92 arcsec, <$\mu_e$> = 17.68 and $M_J=-23.26$. We adopted a colour {\sl g--J} = 2.86 \citep{mannucci01, fukugita96} from which $M_{g} = -20.40$, $\mu_{\rm e} = 20.95$ and $\log R_{\rm e} = 0.396$.

From the 2D modelling and 1D profile, we then conclude that a bulge component is sufficient to account for the host galaxy structure in our image. The BH mass was computed using the relationship provided by \citet{marconi03} for the sub-group of galaxies with a secure BH mass estimate
(i.e. group 1; see their table 2). The absolute $J$-band magnitude of the bulge is provided in Table \ref{ta:galfpar}, and is corrected for the
Galactic extinction. The $K$-correction was found to be negligible according
to \citet{Chilingarian10}. The resulting BH mass is (2.1 $\pm$ 0.2) $\times 10^{8}$\,M$_{\odot}$. Considering the dispersion in the \citet{marconi03} $L_J$--$M_{\rm BH}$ relation, we estimate a factor of $\sim 2$ uncertainty on $M_{\rm BH}$.

\section{Discussion}

In this Letter, we presented a structural modelling of the host galaxy of the
$\gamma$-ray-emitting NLSy1 FBQS J1644$+$2619. The 2D surface brightness is
fitted up to 5 arcsec (corresponding to 13 kpc) by the combination of a
nuclear component and a bulge component with a S\'ersic profile with index $n$
= 3.7, indicative of an elliptical galaxy. The low ellipticity suggests that
the host is an E1 galaxy and the structural parameters are consistent with the
correlations of effective radius and surface brightness with absolute
magnitude measured for elliptical galaxies. Our results seem to contradict the
results reported in \citet{olguin17} for the same target. Studying its
morphology in the $J$ and $K_s$ bands, they claim for a pseudo-bulge nature of the
host. Although our deeper observation traces the host profile to a larger
radius than those reported in \citet{olguin17}, we do not find any evidence of
the presence of a stellar bar that they report in our PSF-subtracted image
residual. However, we cannot rule out the presence of other low-luminosity
components such as, for example, a disc or a bar, on spatial scales larger than about 3 arcsec, where some residuals are present in our fit of the surface brightness distribution of the source (see Fig.~\ref{fig:radprof}, lower panel).

Among the other NLSy1 detected by {\em Fermi}-LAT in $\gamma$-rays up to now, the morphology of the host galaxy has been investigated only for 1H 0323$+$342 and PKS 2004$-$447. Observations of 1H 0323$+$342 with the {\em Hubble Space Telescope} and the NOT revealed a structure that may be interpreted either as a one-armed galaxy \citep{zhou07} or as a circumnuclear
ring \citep{anton08,leontavares14}, suggesting two possibilities: the spiral
arm of the host galaxy or the residual of a galaxy merger, respectively. In
the case of PKS 2004$-$447, near-IR  Very Large Telescope (VLT) observations
suggested that the host may have a pseudo-bulge morphology
\citep{kotilainen16}. This should imply that the relativistic jet in PKS
2004$-$447 is launched from a pseudo-bulge via secular processes, in contrast
to the conjectures proposed for jet production. However, the surface
brightness distribution of the host is not well constrained by a bulge$+$disc
model at large radii, leaving the debate on its morphology open.

From the $J$-band bulge luminosity of the FBQS J1644$+$2619 host galaxy, we
inferred a BH mass of (2.1 $\pm$ 0.2) $\times$10$^{8}$\,M$_{\odot}$. Earlier
estimates span more than one order of magnitude. \citet{yuan08} calculated a
virial mass $M_{\rm BH}$ = 8 $\times$ 10$^6$ M$_\odot$ from  its optical
spectrum and the broad-line region (BLR) radius--luminosity relation by
\citet{kaspi05}\footnote{We note  that when using the BLR radius--luminosity
  relation from \citet{bentz13} and the BH mass scaling relation based on
  FWHM(H$\beta$) and L(H$\beta$) from  \cite{vestergaard06}, the BH mass would
  increase to 1.8 $\times$ 10$^7$ M$_\odot$.}. By modelling the optical--UV
spectrum with a Shakura \& Sunyaev accretion disc model, \citet{calderone13}
found  $M_{\rm BH}$ = 1.6 $\times$ 10$^{8}$ M$_{\odot}$. This value is compatible with that obtained from our IR observations.

Black hole masses in the range (1.5--2.2) $\times$ 10$^7$ M$_\odot$ were estimated for 1H 0323$+$342 from optical spectroscopy \citep{landt16}, while
values of (1.6--4.0) $\times$ 10$^8$ M$_\odot$ were obtained by using the relation between the BH mass and the bulge luminosity, depending on the model used to reproduce the surface brightness profile \citep{leontavares14}. A significant discrepancy was also found in the BH mass estimates of PKS 2004$-$447. From the $K$-band bulge luminosity and the $M_{\rm\,BH}$--$L_{\rm\,bulge}$ relation for pseudo-bulges from \citet{ho14},
\citet{kotilainen16}  estimated $M_{\rm BH}$ = 9 $\times$ 10$^{7}$ M$_{\odot}$, a value lower than that derived from VLT spectropolarimetric observations of the source \citep[$6\times 10^{8}$ M$_{\odot}$;][]{baldi16}. However, the mass of the host galaxy inferred from the $K$-band luminosity by \citet{kotilainen16}, $M_{\rm tot}$ = 7 $\times$ 10$^{11}$ M$_{\odot}$, is a typical value of a giant elliptical galaxy. From that value, by using the relation between BH mass, bulge mass, and near-IR luminosity proposed by \cite{marconi03}, we obtain a BH mass of 10$^9$ M$_\odot$, not compatible with a spiral galaxy. 

As shown earlier, there is a discrepancy between the values of the BH mass obtained from the bulge luminosity and those obtained by scaling relations based on the emission-line properties. \citet{leontavares14}  proposed an explanation in which the BLR has a flat structure and the line profile depends on the inclination of its axis with respect to the line of sight. A similar scenario was proposed by \citet{baldi16}. Assuming that $\gamma$-ray emitters are oriented typically at $i \le 5^\circ$, the mass value obtained from the line profiles must be corrected by adding a factor between 0.84 and 1.16 dex \citep{Decarli08}, resulting in $M_{\rm\,BH} \ge 10^{8}$ M$_\odot$, which brings in better agreement the two types of BH mass estimations. 

In view of the results of this Letter and of the above discussion, we conclude
that there is no clear evidence that the $\gamma$-ray-emitting NLSy1 are
hosted in spiral galaxies with low BH masses. In contrast, in the case of FBQS J1644$+$2619 the host is a typical E1 galaxy and the BH mass is well above the limit characterizing radio-loud AGN, i.e.\ $\log M_{\rm BH}$/M$_\odot \sim 7.8$ \citep{baldi10}.

\section{Summary}

Several studies indicate that powerful jets in AGN are produced only by the most massive black holes, with $M_{\rm BH} \ga 10^8$ M$_\odot$. This idea has been challenged by the discovery of $\gamma$-ray emission from a few radio-loud NLSy1 galaxies, usually hosted in spiral galaxies with small BH masses ($M_{\rm BH} \sim 10^6$--$10^7$ M$_\odot$), suggesting that their relativistic jets might be produced by a different mechanism. However, it has been claimed that the low $M_{\rm BH}$ estimates might be due to projection and/or radiation pressure effects, while the optical classification of the galaxy hosting these objects is not clear.

In this Letter, we presented near-IR data of the $\gamma$-ray-emitting NLSy1
galaxy FBQS J1644$+$2619, which were collected in the $J$ band at the GTC with
CIRCE. The aim was to establish the morphology of its host galaxy and to make
a reliable estimate of its BH mass. The 2D surface brightness profile of the
host galaxy of FBQS J1644$+$2619 is well described by an unresolved nuclear
component and a single bulge component with a S\'ersic profile with index $n$
= 3.7, assessing an early-type galaxy profile.  The BH mass estimated by the
IR bulge luminosity is (2.1 $\pm$ 0.2) $\times$ 10$^{8}$ M$_{\odot}$. All these pieces of evidence indicate that the relativistic jet in this NLSy1 is produced by a massive BH hosted in an
elliptical galaxy with $M_{\rm BH} \ga 10^8$ M$_\odot$, in agreement with the other radio-loud AGN. This is a key issue in the context of our understanding of the production of powerful relativistic jets in radio-loud AGN. Since the literature shows conflicting results, further high-resolution observations of the host galaxy of $\gamma$-ray-emitting NLSy1 are needed to cast more light on the sub-population of very radio loud NLSy1 and clarify the nature of their hosts.

\section*{Acknowledgements}
This Letter is based on observations made with the GTC telescope, in the Spanish Observatorio del Roque de los Muchachos of the Instituto de
Astrofisica de Canarias, under Director's Discretionary Time (proposal code GTC2016-053). CRA acknowledges the Ram\'on y Cajal Program of the Spanish Ministry of Economy and Competitiveness (RYC-2014-15779). Development of CIRCE was supported by the University of Florida and the National Science Foundation (grant AST-0352664), in collaboration with IUCAA.

\bsp	
\label{lastpage}

\begin{thebibliography}{99}
\bibitem[Abdo et al.(2009)]{abdo09} Abdo A. A. et al., 2009, ApJ, 707, L142   
\bibitem[Anton et al.(2008)]{anton08} Anton S., Browne I. W., Marcha M. J., 2008, A\&A, 490, 583 
\bibitem[Bade et al.(1995)]{bade95} Bade N., Fink H. H., Engels D., Voges W., Hagen H.-J., Wisotzki L., Reimers D., 1995, A\&AS, 110, 469
\bibitem[Baldi \& Capetti(2010)]{baldi10} Baldi R., Capetti A., 2010, A\&A, 519, A48
\bibitem[Baldi et al.(2016)]{baldi16} Baldi R., Capetti A., Robinson A., Laor A., Behar E., 2016, MNRAS, 458, L69                        
\bibitem[Bentz et al.(2013)]{bentz13} Bentz M. C. et al., 2013, ApJ, 767, 149
\bibitem[Bernardi et al.(2003)]{bernardi03} Bernardi M. et al., 2003, AJ, 125, 1849
\bibitem[Blanton et al.(2003)]{blanton03} Blanton M. R. et al., 2003, ApJ, 594, 186
\bibitem[B\"ottcher $\&$ Dermer(2002)]{boettcher02} B\"ottcher M., Dermer C. D., 2002, ApJ, 564, 86  
\bibitem[Calderone et al.(2013)]{calderone13} Calderone G., Ghisellini G., Colpi M., Dotti M., 2013, MNRAS, 431, 210                 
\bibitem[Capetti \& Balmaverde(2006)]{capetti06} Capetti A., Balmaverde B., 2006, A\&A, 453, 27            
\bibitem[Chiaberge et al.(2015)]{chiaberge15} Chiaberge M., Gilli R., Lotz J. M., Norman C., 2015, ApJ, 806, 147 
\bibitem[Chilingarian, Melchior \& Zolotukhin(2010)]{Chilingarian10} Chilingarian I. V., Melchior A.-L., Zolotukhin I. Y., 2010, MNRAS, 405, 1409 
\bibitem[D'Ammando et al.(2012)]{dammando12} D'Ammando F. et al., 2012, MNRAS, 426, 317                      
\bibitem[D'Ammando et al.(2015)]{dammando15} D'Ammando F., Orienti M., Larsson J., Giroletti M., 2015, MNRAS, 452, 520 
\bibitem[D'Ammando et al.(2016)]{dammando16} D'Ammando F., Orienti M., Finke J., Larsson J., Giroletti M., Raiteri C., 2016, Galaxies, 4, 11  
\bibitem[Decarli et al.(2008)]{Decarli08} Decarli R., Dotti M., Fontana M., Haardt F., 2008, MNRAS, 386, L15
\bibitem[Deo et al.(2006)]{deo06} Deo R. P., Crenshaw D. M., Kraemer S. B., 2006, AJ, 132, 321 
\bibitem[Fukugita et al.(2006)]{fukugita96} Fukugita M., Ichikawa T., Gunn J.~E., Doi M., Shimasaku K., Schneider D.~P., 2006, AJ, 111, 1748
\bibitem[Garner et al.(2014)]{Garner14} Garner A. et al., 2014, in Ramsay S. K., McLean I. S., Takami H., eds, Proc. SPIE Conf. Ser. Vol. 9147, Ground-Based and Airborne Instrumentation for Astronomy V. SPIE, Bellingham, p. 91474A
\bibitem[Ho \& Kim(2014)]{ho14} Ho L. C., Kim M., 2014, ApJ, 789, 17
\bibitem[Hopkins et al.(2008)]{hopkins08} Hopkins P. F., Hernquist L., Cox T. J., Keres D., 2008, ApJS, 175, 356
\bibitem[Kaspi et al.(2005)]{kaspi05} Kaspi S., Maoz D., Netzer H., Peterson B. M., Vestergaard M., Jannuzi B. T., 2005, ApJ, 629, 61
\bibitem[Komossa et al.(2006)]{komossa06} Komossa S., Voges W., Xu D., Mathur S., Adorf H.-M., Lemson G., Duschl W. J., Grupe D., 2006, AJ, 132, 531
\bibitem[Kotilainen et al.(2016)]{kotilainen16} Kotilainen J. K., Leon-Tavares J., Olguin-Iglesias A., Baes M., Anorve C., Chavushyan V., Carrasco L., 2016, ApJ, 832, 157 
\bibitem[Leon-Tavares et al.(2014)]{leontavares14} Leon Tavares J. et al., 2014, ApJ, 795, 58 
\bibitem[Landt et al.(2017)]{landt16} Landt H. et al., 2017, MNRAS, 464, 2565
\bibitem[Lister et al.(2016)]{lister16} Lister M. L. et al., 2016, AJ, 152, 12
\bibitem[Mannucci et al.(2001)]{mannucci01} Mannucci F., Basile F., Poggianti B.~M., Cimatti A., Daddi E., Pozzetti L., Vanzi L., 2001, MNRAS, 326, 745
\bibitem[Marconi \& Hunt(2003)]{marconi03} Marconi A., Hunt L., 2003, ApJ, 589, L21  
\bibitem[Marconi et al.(2008)]{marconi08} Marconi A., Axon D. J., Maiolino R., Nagao T., Pastorini G., Pietrini P., Robinson A., Torricelli G., 2008, ApJ, 678, 693 
\bibitem[Markarian et al.(1989)]{markarian89} Markarian B. E., Lipovetsky V. A., Stepanian J. A., Erastova L. K., Shapovalova A. I., 1989, Soobshch. Spets. Astrofiz. Obs., 62, 5
\bibitem[Marscher(2010)]{marscher10} Marscher A., 2010, in Belloni T., ed., Lecture Notes in Physics, Vol. 794, The Jet Paradigm, Springer-Verlag, Berlin, p. 173 
\bibitem[Mathur et al.(2000)]{mathur00} Mathur S., 2000, MNRAS, 314, 17
\bibitem[Mathur et al.(2012)]{mathur12} Mathur S., Fields D., Peterson B. M., Grupe D., 2012, ApJ, 754, 146
\bibitem[Morganti et al.(2011)]{morganti11} Morganti R., Holt J., Tadhunter C., Ramos Almeida C., Dicken D., Inskip K., Oosterloo T., Tzioumis T., 2011, A\&A, 535A, 97
\bibitem[Olguin-Iglesias et al.(2017)]{olguin17} Olguin-Iglesias A., Kotilainen J. K., Leon Tavares J., Chavushyan V., Anorveet C., 2017, MNRAS, 467, 3712
\bibitem[Osterbrock $\&$ Pogge(1985)]{osterbrock85} Osterbrock D. E., Pogge R. W., 1985, ApJ, 297, 166
\bibitem[Peng et al.(2002)]{Peng02} Peng C. Y., Ho L. C., Impey C. D., Rix H.-W., 2002, AJ, 124, 266
\bibitem[Pogge(2000)]{pogge00} Pogge R. W., 2000, New Astron. Rev., 44, 381
\bibitem[Sikora, Stawarz \& Lasota(2007)]{sikora07} Sikora M., Stawarz L., Lasota J.-P., 2007, ApJ, 658, 815 
\bibitem[Singh et al.(2015)]{singh15} Singh V., Ishwara-Chandra C. H., Sievers J., Wadadekar Y., Hilton M., Beelen A., 2015, MNRAS, 454, 1556
\bibitem[Vestergaard \& Peterson(2006)]{vestergaard06} Vestergaard M., Peterson B. M., 2006, ApJ, 641, 689
\bibitem[Yuan et al.(2008)]{yuan08} Yuan W., Zhou H. Y., Komossa S., Dong X. B., Wang T. G., Lu H. L., Bai J. M., 2008, ApJ, 685, 801      
\bibitem[Woo \& Urry(2002)]{woo02} Woo J.-H., Urry M., 2002, ApJ, 579, 530 
\bibitem[Zhou et al.(2007)]{zhou07} Zhou H. et al., 2007, ApJ, 658, L13
\end{thebibliography}
\end{document}